\documentclass{mem}
\usepackage{natbib}\usepackage{txfonts}\usepackage{balance}
\usepackage{graphicx}
\usepackage[a4paper,breaklinks,dvipdfm]{hyperref}
\idline{75}{282}
\begin{document}

\title{
A new nearby PWN overlapping the Vela Jr SNR
}

%   \subtitle{}

\author{
F. \,Acero\inst{1}, 
Y. \, Gallant\inst{1},
R. \, Terrier\inst{2},
M. \, Renaud\inst{1}
\and J. \, Ballet\inst{3}
          }

  \offprints{F. Acero}

\institute{
LUPM, CNRS/Universit\'e Montpellier 2
\email{facero@in2p3.fr}
\and
APC, CNRS/Universit\'e Paris Diderot-Paris 7 
\and 
AIM, CEA/CNRS/Universit\'e Paris Diderot-Paris 7 
}

\authorrunning{Acero }

\titlerunning{Discovery of a nebula around the nearby pulsar PSR\,J0855$-$4644 }

\abstract{

PSR\,J0855$-$4644 is an energetic pulsar ($\dot{E} = 1.1 \times 10^{36}$ erg/s, $P$=65 ms) discovered near the South-East rim of the supernova remnant (SNR) RX\,J0852.0$-$4622 (aka Vela Jr) by the Parkes Multibeam Survey. The position of the pulsar is in spatial coincidence with an enhancement in X-rays and TeV $\gamma$-rays, which could represent its pulsar wind nebula (PWN). 

We have revealed with an XMM-Newton observation the X-ray counterpart of the pulsar together with a surrounding extended emission thus confirming the suggestion of a PWN. The comparison of the absorption column density derived in X-rays from the pulsar with $^{12}$CO observations (tracing the dense gas) is used to derive an upper limit to the distance of the pulsar (d$<$ 900 pc) and to discuss a possible association 
of the pulsar with the Vela Jr SNR. 
This new distance estimate implies that the pulsar is nearby and could therefore significantly contribute to the observed spectrum of
cosmic-ray leptons (e$^{-}$/e$^{+}$). 


}
\maketitle{}

\section{Introduction}

PSR\,J0855$-$4644  is a young and energetic pulsar (characteristic age $\tau_{\rm c}$=140 kyr, $\dot{E} = 1.1 \times 10^{36}$ erg/s) with a period of 65 ms recently
discovered in the Parkes multibeam survey \citep{kramer03} lying on the South-East rim of the Vela Jr SNR.

The position of the pulsar is in spatial coincidence with an enhancement in the X- and $\gamma$-ray maps
 \citep[presented in ][ respectively]{aschenbach99,ah07-Vela Jr}, which could represent its pulsar wind nebula. 

The distance to the pulsar is estimated to be 4 kpc based on the radio dispersion measure combined with 
the NE2001 model of Galactic electron distribution \citep{cordes02}.

However, the estimation of the distance using such method in this specific region of the sky can be problematic 
due to the structured and nearby emission of the Gum nebula and the Vela and Vela Jr  SNRs.

In order to better constrain the distance to the pulsar, and search for a potential X-ray nebula, 
an  \textit{XMM-Newton} observation was carried out whose results are presented in this article.

\begin{figure*}[t]
 \centering
% Two side by side figures%
   \begin{tabular}{ccc}
 {\includegraphics[bb= 50 145 530 600,clip,width=4.5cm]{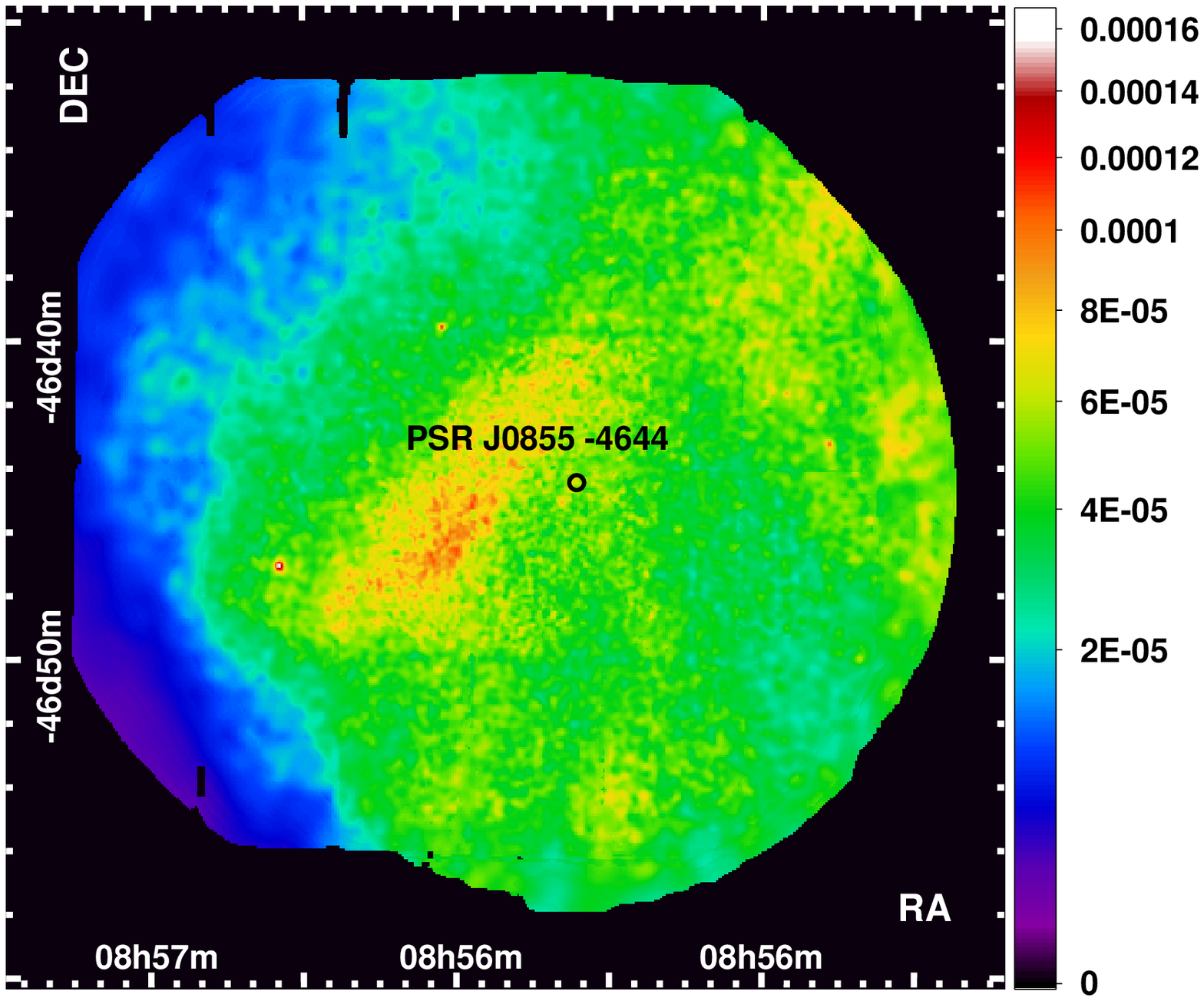} } & \hspace{-6mm}
 {\includegraphics[bb= 40 145 530 600,clip,width=4.55cm]{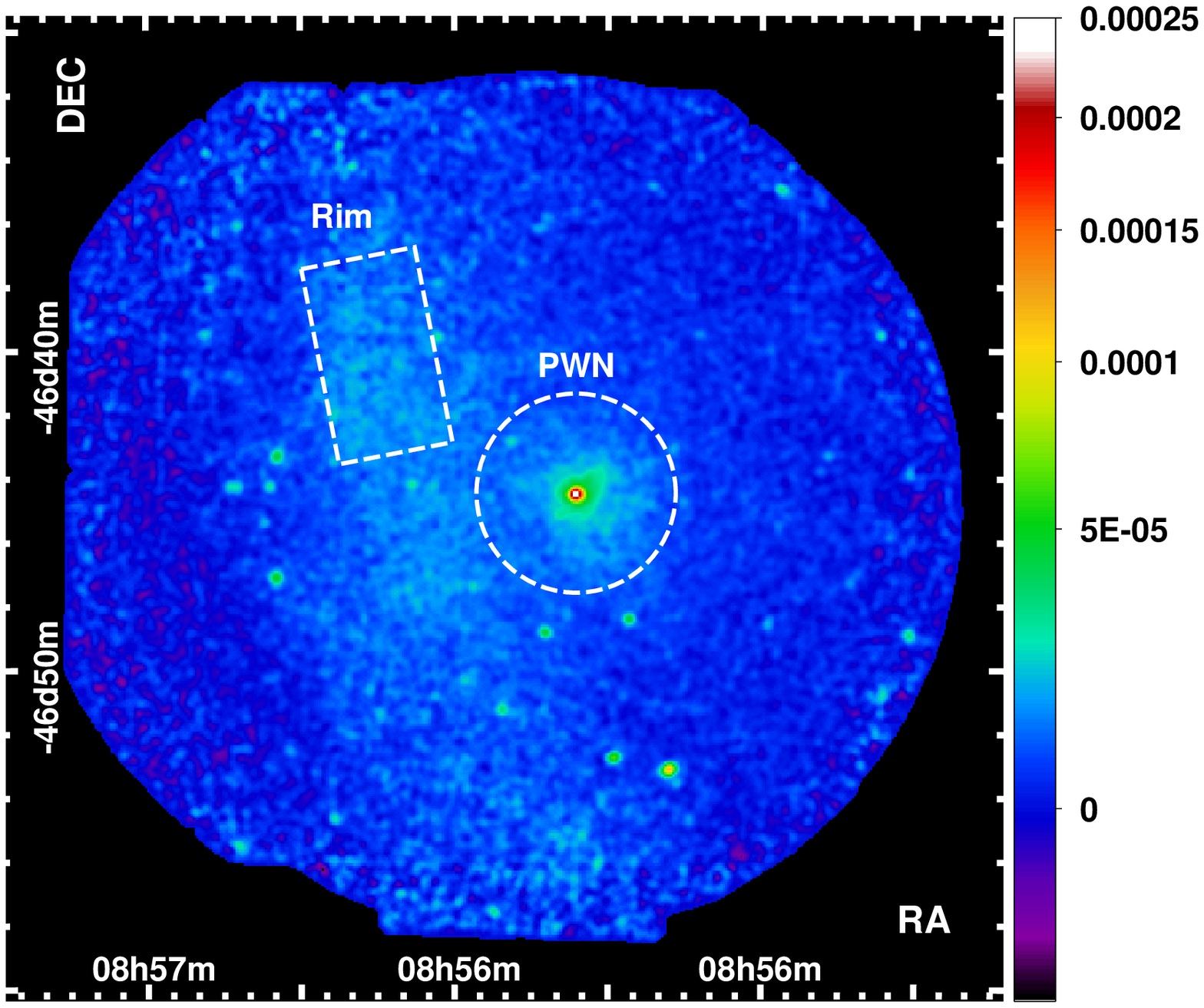} } & \hspace{-6mm}
{\includegraphics[bb= 90 370 495 790,clip,width=4.2cm]{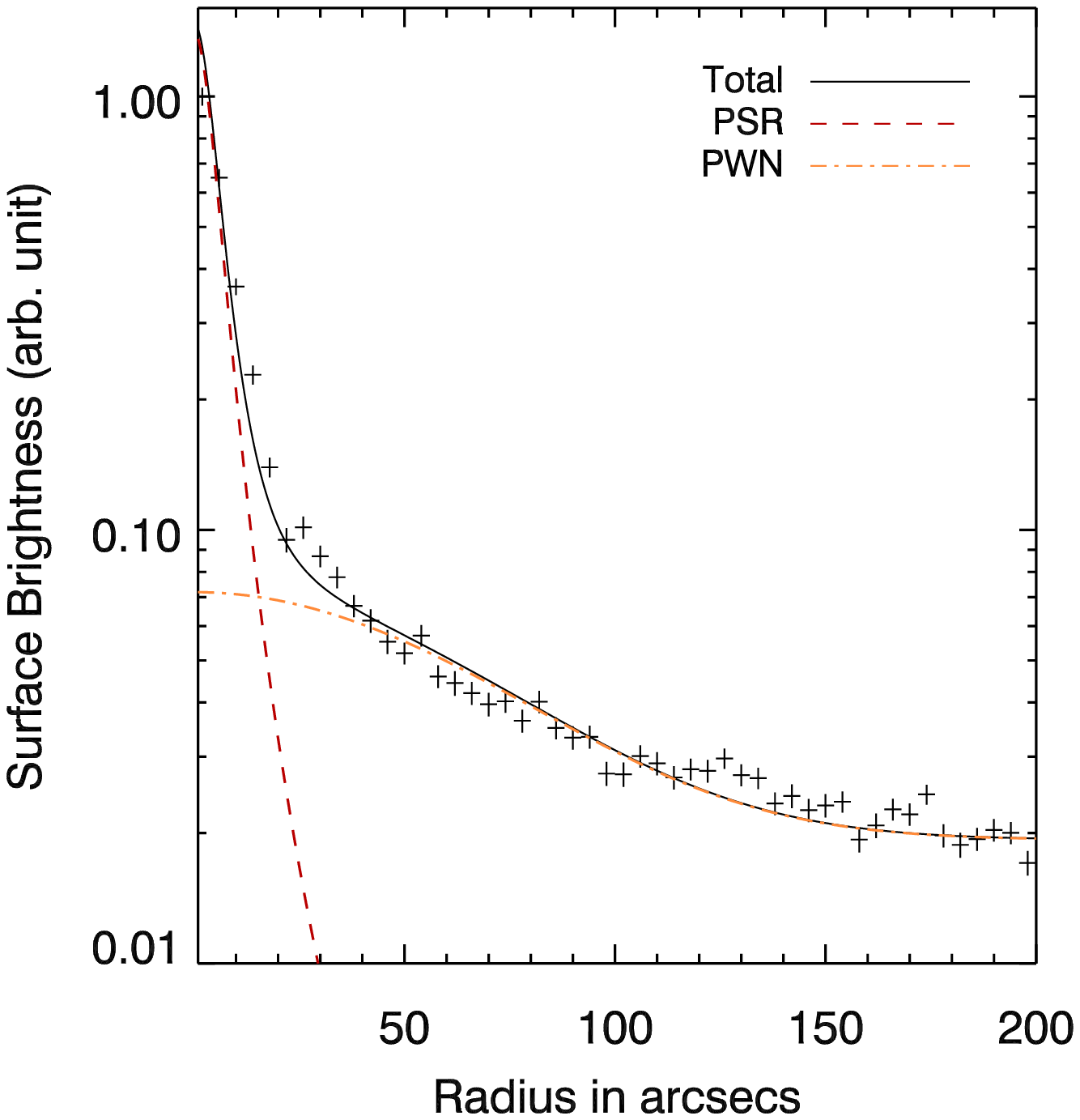} }

   \end{tabular}

\vspace{-3mm}

\caption{
\footnotesize
\textit{XMM-Newton} image combining MOS plus PN cameras with units in ph/cm$^{2}$/s/arcmin$^{2}$. \textit{Left}: low-energy band image (0.5-0.8 keV) 
adaptively smoothed to a signal-to-noise ratio of 10.
 The position of PSR\,J0855$-$4644 is shown with a black circle.  \textit{Middle}: image in the high-energy band (1.2-6 keV) smoothed with a Gaussian width of 7''.
 The box illustrates the region of spectral extraction used to study the rim of the Vela Jr SNR and  the
  circle represents the maximal extent of the radial profile shown in the next panel.
 \textit{Right}: normalized radial profile of the PWN in the high-energy band extracted from a region centered on the pulsar.
 The best-fit two components model (point-like + diffuse) is overlaid.
}
\label{xmm}
\end{figure*}

\section{X-ray observations}
\label{obs}

 \begin{figure}[]
\resizebox{\hsize}{!}{\includegraphics[bb= 70 25 590 705,clip,angle=-90]{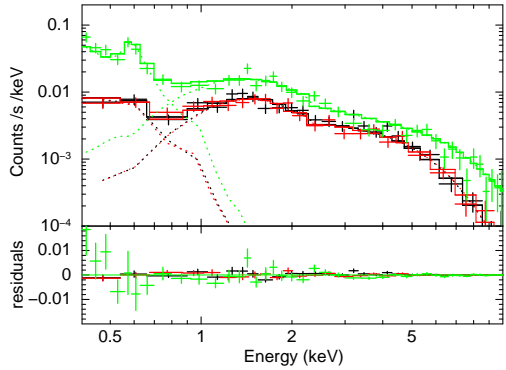}}
\vspace{-0.3cm}
\caption{
\footnotesize
X-ray spectrum of PSR\,J0855$-$4644, using the MOS1, MOS2 and PN camera (in black, red and green respectively), extracted from a 15'' region.
The  thermal emission from the Vela SNR, dominating at low energies, is described by an APEC model (astrophysical plasma emission code)
   while the emission from the pulsar is represented by an absorbed power-law model (dashed line). 
}
\label{spec}
\end{figure}

The pulsar was observed with \textit{XMM-Newton} for 55 ks using the imaging mode for the three EPIC instruments. 
After flare screening, the remaining exposure time is 40 ks and 29 ks for the MOS and PN cameras respectively.
The background subtracted and vignetting corrected images in a low (0.5-0.8 keV) and high-energy band (1.2-6 keV) are presented in Fig. \ref{xmm}.
Those images clearly show the two very different  facets of the Vela Jr region. While the low energies are dominated by the thermal emission
from the large scale Vela SNR, the high energy image reveals a faint rim of the Vela Jr SNR and, for the first time, the X-ray counterpart of the pulsar.
Interestingly, some diffuse emission is also seen surrounding the pulsar. 
To further investigate the morphology of this diffuse emission, a radial profile, centered on the position of the pulsar, was extracted.
The profile was then fitted with a combination of a point-like component (using the \textit{XMM-Newton} PSF from the calibration files)
 plus an extended component (a Gaussian profile) representing the pulsar and the nebula respectively as shown in Fig. \ref{xmm} (right panel).
The best-fit Gaussian width is (57$\pm$2)'' and the diffuse emission around the pulsar is clearly detected up to $\sim$150''.
 The large scale emission is fairly symmetric and no indications of a trail or of a bow-shock structure 
 (reminiscent of a fast moving pulsar) are observed.
 
 In order to derive the X-ray  properties of the pulsar, we have extracted a spectrum in a 15'' region centered on the pulsar.
The thermal emission from Vela was fixed using a template model obtained in an annulus region of $15''<r<50''$ around the pulsar
and renormalized to the ratio of geometrical areas. 
The instrumental and particle induced background was derived from the compilation of observations with the filter wheel 
in closed position and renormalized in the 10-12 keV energy band over the whole  field of view.

\begin{table*}
%\begin{minipage}[t]{\columnwidth}
\centering
\caption{Best-fit parameters obtained on different parts of the SNR's rim and on the pulsar.  
The parameters from the thermal component for the pulsar are frozen while for the other regions, the error on the 
temperature is of the order of 5 eV and the N$_{\rm H}$ is not well constrained.
The unabsorbed non-thermal flux is given in the 2-10 keV energy range. The error bars are given at 90\% confidence level.}
\label{tab}
\begin{tabular}{l | c c c c c  }     % 8 columns %
\hline

 &   \multicolumn{2}{c}{Thermal component}  & \multicolumn{3}{c}{Non-thermal component}     \\
Parameters & N$_{\rm H}$ (10$^{22}$ cm$^{-2}$)& $kT$ (keV) & N$_{\rm H} $ & Index & Flux (10$^{-12}$ ergs/cm$^{2}$/s) \\

\hline
  \rule[-7pt]{0pt}{17pt} 
   PSR\,J0855$-$4644 & $5 \times 10^{-2} $ & 0.14 &  0.62$^{+0.07}_{-0.10}$ & 1.30$^{+0.09}_{-0.09}$ & 0.30$^{+0.02}_{-0.05}$   \\
   
   \rule[-7pt]{0pt}{17pt} 
   South-East rim & $2 \times 10^{-2}$ & 0.15 & 0.87$^{+0.08}_{-0.05}$ & 2.38$^{+0.03}_{-0.03}$ & 1.83$^{+0.08}_{-0.09}$  \\
   
     \rule[-7pt]{0pt}{17pt} 
   South rim &  $<1 \times 10^{-2} $   & 0.18 &  0.82$^{+0.06}_{-0.05}$ & 2.68$^{+0.05}_{-0.06}$ &   3.35$^{+0.09}_{-0.08}$ \\
   
     \rule[-7pt]{0pt}{17pt} 
   North-West rim & $<2 \times 10^{-2} $ & 0.17   & 0.67$^{+0.04}_{-0.06}$ & 2.55$^{+0.05}_{-0.04}$ & 3.85$^{+0.11}_{-0.13}$ \\
   
\hline
\end{tabular}
%\end{minipage}
\end{table*}

The spectrum of the pulsar (presented in Fig. \ref{spec}) is well described at low energies by the fixed template
 (absorbed APEC model) and at high energies with an absorbed power-law model. 
To compare the X-ray absorption column towards other parts of the rim of Vela Jr SNR, we
have applied the same spectral analysis method to the south-east rim in the same pointing and to two other \textit{XMM-Newton} observations whose pointing positions are  shown in Fig. \ref{CO} (South and North-West pointings).
The resulting  best-fit parameters of the spectral analysis discussed in this section are listed in Table \ref{tab}.

\section{Distance estimate}
\label{distance}

\begin{figure}[]
\resizebox{\hsize}{!}{\includegraphics[bb= 45 150 570 650,clip]{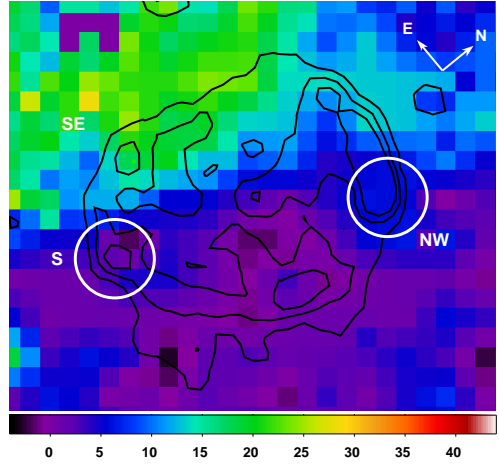}}
\vspace{-0.4cm}
\caption{
\footnotesize
$^{\rm 12}$CO map in the direction of the Vela Jr SNR integrated on the whole LSR velocity range. The linear scale is in units of K km s$^{-1}$.
The contours of the ROSAT X-ray emission from the SNR are shown in black. The \textit{XMM-Newton} observations used in this article are shown with the circles and 
the position of PSR\,J0855$-$4644 is indicated by the white cross.
}
\label{CO}
\end{figure}

The method described in this paragraph to  tackle the issue of the distance to the pulsar
 is based on  the existence of a strong contrast of integrated $^{12}$CO \citep[using the survey from][]{dame01},  between the eastern 
 and the western part of the SNR as shown in Fig. \ref{CO}.
This contrast is due to the Vela Molecular Ridge (VMR) cloud C \citep[see ][ for clouds nomenclature]{murphy91},
 lying at a distance of 700$\pm$200 pc \citep{liseau92}, which is the main contributor of $^{12}$CO  along the line of sight.
 Therefore if the Vela Jr SNR lies in the background of the VMR, the observed gradient in $^{12}$CO should also be seen
 in the X-ray absorption column.
 
\begin{figure}[]
\resizebox{\hsize}{!}{\includegraphics[bb= 85 370 550 700,clip]{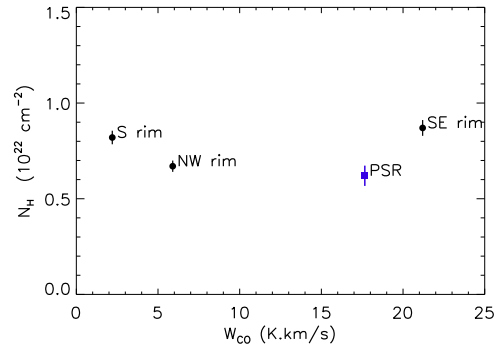}}
\vspace{-0.4cm}
\caption{
\footnotesize
Correlation plot between the integrated $^{12}$CO value on the line of sight and the best-fit X-ray absorption column derived from \textit{XMM-Newton} observations.
The X-ray column derived from the rims of the Vela Jr SNR and from PSR\,J0855$-$4644 are represented by black circles and a blue square respectively.
The label over each point corresponds to regions where the values have been measured as defined in Fig. \ref{CO}.}
\label{correl}
\end{figure}

This hypothesis is tested in the correlation plot in Fig. \ref{correl}, where the value of the velocity integrated $^{12}$CO (W$_{\rm CO}$) for each 
position in the sky is plotted against the best-fit value of the X-ray absorption on the pulsar and on the other regions of the SNR's rim.
No correlation is observed  in such a plot and we note that the $N_{\rm H}$ on the pulsar (0.62$^{+0.07}_{-0.10}\times 10^{22}$ cm$^{-2}$) 
is slightly less than the column on the SNR's rim in the same region (0.87$^{+0.08}_{-0.05}\times 10^{22}$ cm$^{-2}$).

We therefore conclude that both the SNR and the pulsar are lying in the foreground of the VMR i.e. at a distance d $<$ 900 pc with the 
pulsar being preferentially in front of the SNR.
This distance estimate of the Vela Jr SNR is compatible with a recent estimate based on the measurement of the proper motion
of the shock in X-rays in the North-western part of the rim by \citet{katsuda10} who derived d$\sim$750 pc.
In addition, the closer distance of the pulsar (instead of 4 kpc) confirms a suggestion made by \citet{redman05} indicating
a distance of $\sim$750 pc. 

Although the pulsar and the SNR may be at similar distances, an association is unlikely as the required pulsar's kick velocity to
explain its current position would be $\sim$3000 km/s in order to travel 12 pc (1$^{\circ}$ at 750 pc) for a SNR age of 4000 years,  much higher than observed in other pulsars \citep{hobbs05}. Moreover for such a high shock speed a bow-shock or a tail structure should appear which is not the case.

This upper limit to the distance of the pulsar is very different from the one estimated with the radio dispersion
measure (d=4 kpc) and therefore implies that PSR\,J0855$-$4644 is an energetic and nearby pulsar. After the Vela pulsar, it is the most energetic
pulsar within 1 kpc and could therefore significantly contribute to the spectrum of cosmic-rays e$^{-}$/e$^{+}$  received at Earth  \citep[see][for the contribution of nearby pulsars]{delahaye10}.

\section{Conclusions}

The search for the X-ray counterpart of the PSR\,J0855$-$4644 using \textit{XMM-Newton} led to the following conclusions:

\begin{enumerate}

\item  The X-ray counterpart of the pulsar has been discovered and a surrounding extended emission of non-thermal origin was found.
 The nebula of the pulsar is fairly symmetric and has a maximum extent of the order of 150''.

\item A faint rim of non-thermal emission representing the South-eastern part of the Vela Jr SNR was detected. 
No bright thin X-ray filaments were found as it is observed in the North-West of the SNR.

\item By comparing  the X-ray column density on the pulsar and 
on different regions of the SNR with the dense gas map (using $^{12}$CO emission as a tracer), we 
derived an upper limit to the distance of the SNR and the pulsar: d$<$ 900 pc.

\end{enumerate}

\bibliographystyle{aa}

\end{document}